\documentclass[conference]{IEEEtran}

\usepackage{epsfig}
\usepackage{times}
\usepackage{float}
\usepackage{afterpage}
\usepackage{amsmath}
\usepackage{amstext,cite}
\usepackage{amssymb,bm}
\usepackage{latexsym}
\usepackage{color}
\usepackage{graphicx}
\usepackage{amsmath}
\usepackage{amsthm}
\usepackage{graphicx}
\usepackage[center]{caption}
\usepackage{pstricks}
\usepackage{caption}
\usepackage{subcaption}
\usepackage{booktabs}
\usepackage{multicol}
\usepackage{lipsum}
\usepackage[T1]{fontenc}
\usepackage{aecompl}
 \usepackage{mathrsfs}
\allowdisplaybreaks

\long\def\comment#1{}

\newcommand{\dv}{{\mathbf d}}

\newcommand{\pv}{{\mathbf p}}
\newcommand{\qv}{{\mathbf q}}

\newcommand{\Cc}{{\mathcal C}}

\newcommand{\Lc}{{\mathcal L}}

\newcommand{\Sc}{{\mathcal S}}

\newcommand{\Wc}{{\mathcal W}}

\newcommand{\Bsf}{{\mathsf B}}

\newcommand{\Ksf}{{\mathsf K}}

\newcommand{\Msf}{{\mathsf M}}
\newcommand{\Nsf}{{\mathsf N}}

\newcommand{\Rsf}{{\mathsf R}}

\newcommand{\Usf}{{\mathsf U}}

\newtheorem{thm}{Theorem}

\newtheorem{rem}{Remark}

\providecommand{\definitionname}{Definition}

\begin{document}

\title{Device-to-Device  Private Caching with Trusted Server}

\author{
\IEEEauthorblockN{%
Kai~Wan\IEEEauthorrefmark{1},
Hua~Sun\IEEEauthorrefmark{2},
Mingyue~Ji\IEEEauthorrefmark{3},
Daniela~Tuninetti\IEEEauthorrefmark{4},
Giuseppe~Caire\IEEEauthorrefmark{1}
}
\IEEEauthorblockA{\IEEEauthorrefmark{1}Technische Universit\"at Berlin, 10587 Berlin, Germany, \{kai.wan, caire\}@tu-berlin.de}%
\IEEEauthorblockA{\IEEEauthorrefmark{2}University of North Texas, Denton, TX 76203, USA, hua.sun@unt.edu}%
\IEEEauthorblockA{\IEEEauthorrefmark{3}University of Utah, Salt Lake City, UT 84112, USA, mingyue.ji@utah.edu}%
\IEEEauthorblockA{\IEEEauthorrefmark{4}University of Illinois at Chicago, Chicago, IL 60607, USA, danielat@uic.edu}%
}

\maketitle

\begin{abstract}
In order to preserve the privacy of   the users demands from other users, in this paper  we formulate a novel information theoretic Device-to-Device (D2D) private caching model by adding a trusted server. In the delivery phase, the trusted server collects the users demands  and sends a query to each user, who then broadcasts packets according to this query. Two D2D private caching schemes (uncoded and coded) are proposed in this paper, which are shown to be order optimal.
\end{abstract}

\section{Introduction}
\label{sec:intro}

Coded caching was originally proposed by Maddah-Ali and Niesen (MAN) for shared-link networks~\cite{dvbt2fundamental}, where a server with access to a library of $\Nsf$ files is connected to $\Ksf$ users  through an error-free broadcast link.
Each user can store up to $\Msf$ files at its cache.  The MAN caching scheme includes {\it placement}  and {\it delivery} phases. In the placement phase without knowing the later demands, letting $t=\Ksf\Msf/\Nsf \in [0:\Ksf]$ represent the ratio between the size of the aggregate cache memory of the $\Ksf$ users and the library size,  
each file is divided into $\binom{\Ksf}{t}$ subfiles, each of which is cached by a different $t$-subset of users. In the delivery phase, each user demands one file. According to the users  demands, the server sends $\binom{\Ksf}{t+1}$ MAN multicast messages, each of which is useful to $t+1$ users simultaneously (i.e., the coded caching/multicasting gain is $t+1$). It was proved in~\cite{ontheoptimality} that the worst-case load achieved by the MAN scheme among all possible demands is optimal under the constraint of uncoded placement (i.e., each user directly stores packets from the library files, rather than more general functions thereof) and $\Nsf \geq \Ksf$.  When $\Nsf\geq \Ksf$, the MAN scheme was also proved in~\cite{yas2} to be generally order optimal within a factor of $2$.
By observing that some MAN  multicast messages when $\Nsf<\Ksf$ are redundant,
	an improved delivery scheme was proposed in~\cite{exactrateuncoded}, which was proved to be   optimal 
under the constraint of uncoded cache placement,  and order optimal within a factor of $2$ without any constraint on the placement; we shall refer
to such a scheme as YMA delivery.

Coded caching strategy was then extended to Device-to-Device networks by Ji, Caire, and Molisch (JCM)~\cite{d2dcaching}, where in the delivery phase  each user broadcasts   packets in function of its cached content and the users demands,  to all other users.
With the MAN cache placement, JCM splits each MAN multicast   message  into $t+1$ equal-length  sub-messages, each of which is conveyed to the other users by a user with
the MAN delivery. By replacing the MAN delivery with the
YMA delivery, the scheme (which effectively splits the D2D network into $K$ parallel shared-link models) is order optimal
to within a factor of $4$ as proved by Yapar et {\it al.} (YWSC)~\cite{ourd2dpaper}.

For the successful decoding of a  MAN multicast message, 
users need to know the composition of this message (i.e., which subfiles are coded together). As a consequence, users are aware of the demands of other users,  which is problematic in terms of privacy.  
Shared-link coded caching with private demands, which aims to preserve the  privacy of the users' demands from other users, was originally discussed in~\cite{Engle2017privatecaching} and recently analyzed information-theoretically  by Wan and Caire (WC) in~\cite{wan2019privatecaching}, where two  schemes were proposed and shown to be order optimal.
  Relevant to this paper is the second scheme in~\cite{wan2019privatecaching} also discussed in~\cite{kamath2019demandpri}, which operates as if there are $\Ksf \Nsf$ users in total, i.e., it pretends there are $\Nsf \Ksf-\Ksf$ virtual users in addition to the $\Ksf$ real users, so that each file is demanded exactly $\Ksf$ times. Such a scheme is order optimal to within a factor of $8$. 
By observing that the private caching schemes in~\cite{wan2019privatecaching,kamath2019demandpri} need high subpacketiation levels (i.e., the number of subfiles into which each file must be partitioned in the placement phase), the authors in~\cite{aravind2019twofiles} proposed a private caching scheme for two-user and two-file systems, with the minimal possible subpacketization level.

\begin{figure}
\centerline{\includegraphics[scale=0.2]{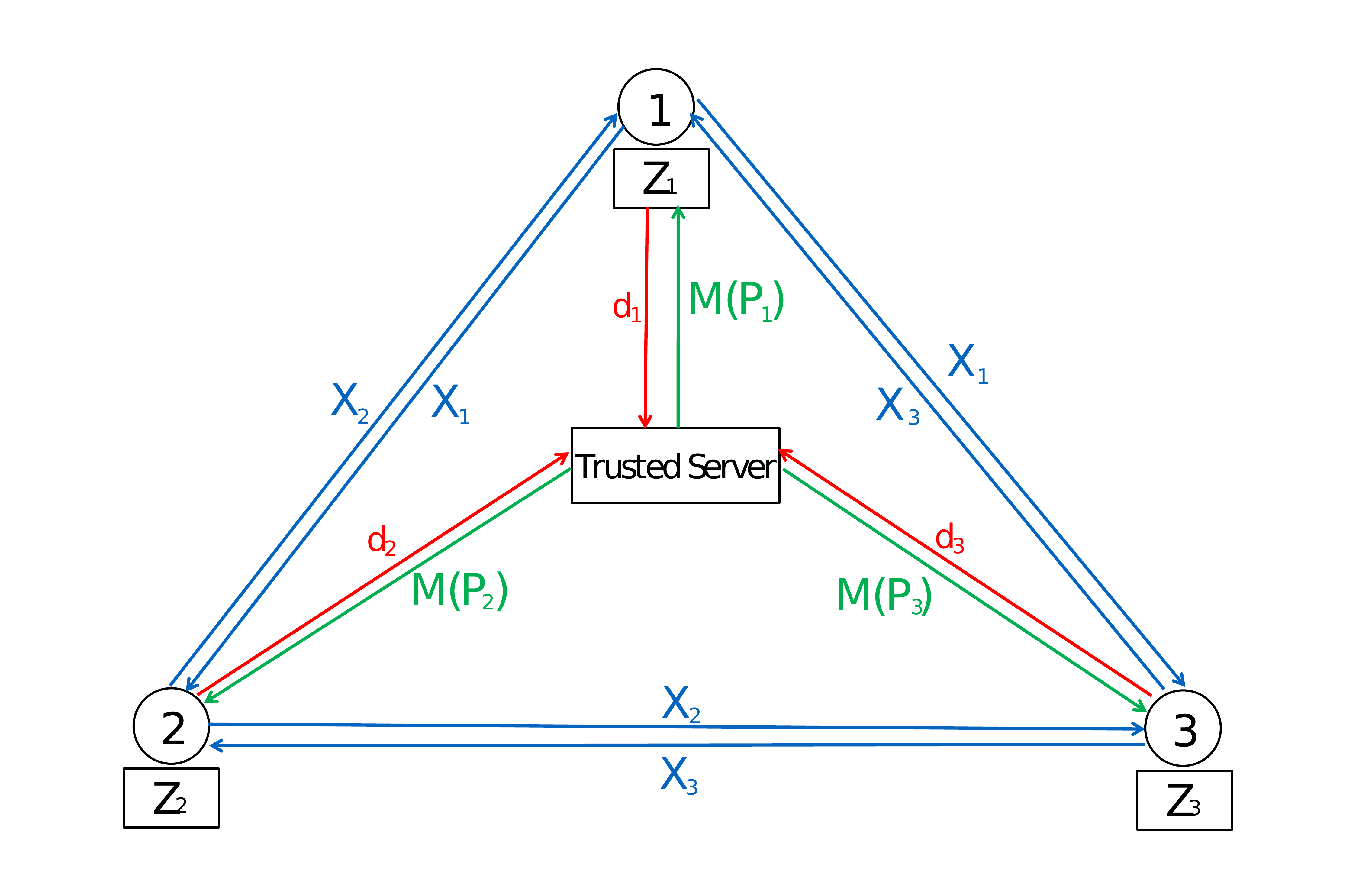}}
\caption{\small The formulated  D2D private caching problem with a trusted server and $\Ksf=3$ users.}
\label{fig: system_model}
\end{figure}

In this paper, we consider the problem of coded caching with private demands for D2D  systems. We introduce a novel  D2D architecture with a {\it trusted server}. This trusted server is connected to each user through an individual link and without access to the library, as illustrated in Fig.~\ref{fig: system_model}. The placement phase is the same as the shared-link and D2D caching models. 
In the delivery phase, each user first informs the trusted server about the index of the demanded file. After collecting the information about the users' demands and the cached content, the trusted server sends a query to each user. Given the query, each user then broadcasts packets accordingly. {\it The trusted server acts only as a coordinator to warrant demand privacy, but does not support any large load of communication.} The demands and the control commands to tell the users what to send can be seen as protocol information, requiring a communication load negligible with respect to the actual file transmission. Hence, the load of the system is still only supported by D2D communication, while the user-server communication is only protocol information. The objective is to design a two-phase private D2D caching scheme with minimum number of transmitted bits by all users in the delivery phase, while preserving the users demands from the other users.



The main contributions in this paper are as follows:
%
(i) We give an information-theoretic formulation of the D2D coded caching problem with demand privacy.
%
(ii) We propose two schemes. 
A baseline uncoded scheme that essentially delivers the whole library to all users, which is trivially private, and a coded scheme that carefully combines the idea of introducing virtual users as in~\cite{wan2019privatecaching,kamath2019demandpri}
with that a splitting the D2D network into multiple shared link ones as in~\cite{d2dcaching,ourd2dpaper}.
%
(iii) By comparing with the  converse bound for the shared-link caching problem without privacy constraint in~\cite{yas2},
we prove the proposed coded scheme is order optimal to within a factor of $6$ when $\Nsf\geq \Ksf$ and $\Msf\geq 2\Nsf/\Ksf$, and within a factor of $12$ when $\Nsf<\Ksf$ and $\Msf\geq \Nsf/\Ksf$.

\paragraph*{Notation Convention}
Calligraphic symbols denote sets, 
bold symbols denote vectors, 
and sans-serif symbols denote system parameters.
We use $|\cdot|$ to represent the cardinality of a set or the length of a vector;
$[a:b]:=\left\{ a,a+1,\ldots,b\right\}$ and $[n] := \{1,2,\ldots,n\}$; 
$\oplus$ represents bit-wise XOR;  
we let $\binom{x}{y}=0$ if $x<0$ or $y<0$ or $x<y$.

 \section{System Model and Related  Results}
\label{sec:model}

\subsection{System Model}
\label{sub:system}

A $(\Ksf,\Nsf,\Msf)$ D2D private caching system with a trusted server
is defined as follows. 
The library contains $\Nsf$ independently generated files, denoted by $(F_{1},F_{2},\dots,F_{\Nsf})$, where each file is composed of $\Bsf$ i.i.d. bits, where 
$\Bsf$ is assumed sufficiently large such that any sub-packetization of the files is possible.
There are $\Ksf$ users in the system, each of which is equipped with a cache of $\Msf \Bsf$ bits, where  $\Msf \in \left[\frac{\Nsf}{\Ksf} ,\Nsf\right]$. 
There is a trusted server without access to the library in the system. This server is connected to each user through an individual  secure  link. In addition, there is also a broadcast link from each user to all other users (e.g., a shared medium)\footnote{We assume a collision avoidance protocol for which when a user broadcasts, all the others stay quiet and listen (e.g., this can be implemented in a practical wireless network using CSMA, as in the IEEE 802.11 standard).}.
We only consider the case $\min\{\Ksf,\Nsf\}\geq 2$, since when $\Ksf=1$ or $\Nsf=1$ each user knows the demand of other users.

 Let $\epsilon_\Bsf \geq 0$ be a constant. 
The system operates in two phases.

{\it Placement Phase.}
Each user $k\in[\Ksf]$ stores content in its cache without knowledge of later demand. 
We denote the content in the cache of user $k\in[\Ksf]$ by 
\begin{align}
Z_{k}=(\mathscr{M}(C_k),C_k),
\label{eq:cache}
\end{align}  
where $C_k$ represents the cached content, a function of the $\Nsf$ files, and $\mathscr{M}(C_k)$ represents the metadata/composition of $C_k$ (i.e., how $C_k$ is generated). We have
\begin{align}
H\big(C_k| \mathscr{M}(C_k),F_{1},\ldots,F_{N} \big)=0 \ \text{(placement constraint)},
\label{eq:cK}
\end{align}  
i.e., $C_k$ is a deterministic function of the library and of the metadata describing the cache encoding. Notice that $\mathscr{M}(C_1),\ldots,\mathscr{M}(C_{\Ksf})$ are random variables over $\Cc_1,\ldots,\Cc_{\Ksf}$, representing all types of cache placement which can be used by the $\Ksf$ users.
In addition, for any $k\in [\Ksf]$, the realization of $\mathscr{M}(C_k)$ is known by user $k$ and the trusted server, and is not   known by other users. The cache content of user $k\in[\Ksf]$ in~\eqref{eq:cache} is constrained by the cache size as
\begin{align}
H(Z_{k}) 
\leq \Bsf(\Msf + \epsilon_\Bsf)  
\ \text{(cache size constraint)}. 
\label{eq:memory size}
\end{align}  

{\it Delivery Phase.} 
During the delivery phase, each user $k\in [\Ksf]$ demands the file 
 with index $d_k$, where $d_k$ is a realization of the random variable $D_k$ with range in $[\Nsf]$.
The demand vector of the $\Ksf$ users, denoted by $\mathbf{D}=(D_1,\ldots,D_{\Ksf})$. 
The delivery phase contains the following steps:
\begin{itemize}

\item Step 1: each user $k\in [\Ksf]$ sends the index of its demanded file (i.e., $d_k$) to the trusted server.

\item Step 2: according to the users' demands and the cache contents, the trusted server 
where the metadata $\mathscr{M}(P_k)$ describes how the packets $P_k$, to be broadcasted by user $k\in [\Ksf]$, are composed.

\item Step 3: each user $k\in [\Ksf]$ broadcasts $X_{k}=(\mathscr{M}(P_k), P_k)$ to other users based only on the its local storage content $Z_k$ and the metadata $\mathscr{M}(P_k)$, that is
\begin{align}
&H(X_{k}|\mathscr{M}(P_k), Z_k )=0  \ \text{(encoding constraint)}.
\label{eq:encoding constraint}
\end{align}
\end{itemize}

{\it Decoding.} 
Let $\mathbf{X} := (X_j: j\in [\Ksf])$ be the vector of all transmitted signals. 
To guarantee successful decoding at user $k\in [\Ksf]$ it must hold that
\begin{align}
 H\big(F_{D_k}| \mathbf{X}, Z_k, D_k \big) \leq \Bsf \epsilon_\Bsf  \ \text{(decoding constraint)},  
\label{eq:decodability}
\end{align}
and to guarantee privacy it must hold 
\begin{align}
 I\big( \mathbf{D}; \mathbf{X}, Z_k |  D_k  \big) =0  \ \text{(privacy  constraint)}.   
\label{eq:privacy2}
\end{align}
The privacy constraint in~\eqref{eq:privacy2} (i.e., vanishing information leakage) corresponds to  perfect  secrecy in an information theoretic sense (see~\cite[Chapter 22]{networkinformation}).

{\it Objective.}
 We say that load $\Rsf$ is achievable if  
\begin{align}
\sum_{k\in[\Ksf]} H(X_{k}) \leq \Bsf(\Rsf+ \epsilon_\Bsf) \ \text{(load)},
\label{eq:loaddef}
\end{align}
while all the above constraints are satisfied and $\lim_{\Bsf\to\infty}\epsilon_\Bsf=0$.
The objective is to determine, for a fixed $\Msf \in \left[\frac{\Nsf}{\Ksf} ,\Nsf\right]$, the minimum achievable load, which is indicated by $\Rsf^{\star}$.

\subsection{Shared-link Private Caching Scheme in~\cite{wan2019privatecaching}}
\label{sub:shared-link private}
We then recall in short the   shared-link private caching scheme proposed in~\cite[Remark 1]{wan2019privatecaching} for general demand and memory size regime, whose key strategy in~\cite{wan2019privatecaching} 
is to generate $\Nsf\Ksf-\Ksf$ virtual users such that the system contains $\Nsf\Ksf$ effective users (i.e., real or virtual users). 


{\it Placement Phase.}
A   private placement precoding strategy was proposed in~\cite{wan2019privatecaching}, which can concatenate with any uncoded cache placement and any MDS-code based placement. 

Now we use this private placement precoding
with the   MAN cache placement for $\Nsf\Ksf$ users. More precisely, let  $\Msf=\Nsf t/(\Nsf \Ksf)=t/\Ksf$ where $t\in [0:\Nsf\Ksf]$.  Each file $F_i$ where $i\in[\Nsf]$ is divided  into $\binom{\Nsf\Ksf}{t}$ non-overlapping and equal-length pieces.
For each file $F_i$, by randomly generating a permutation of $\left[\binom{\Nsf\Ksf}{t}\right]$, we assign each piece to one subfile $F_{i,\Wc}$, where $\Wc\subseteq [\Nsf\Ksf]$ and $|\Wc|=t$, according to this permutation. Each user $k\in [\Ksf]$ caches $F_{i,\Wc}$ where $k\in \Wc$.
The random permutation is unknown by each user $k\in [\Ksf]$. As a result, from the viewpoint of user $k\in [\Ksf]$, each cached subfile of file $F_i$ where  $i\in [\Nsf]$ is equivalent from the viewpoint of user $k$, while each uncached subfile of $F_i$ is also equivalent.

 {\it Delivery Phase.}
 When the demand vector of the $\Ksf$ real users is revealed to the server, the demands of the virtual users are generated such that each file is demanded by exactly $\Ksf$ effective users.
For each $\Sc\subseteq [\Nsf\Ksf]$ where $|\Sc|=t+1$, the server generates a MAN multicast message
\begin{align}
W_{\Sc}=\underset{k\in \Sc}{\oplus } F_{ d_{k}, \Sc\setminus \{k\}}. \label{eq:MAN delivery}
\end{align} 
Then the server, by generating a random permutation of $\left[ \binom{\Nsf\Ksf}{t+1} \right]$,  transmits all $\binom{\Nsf\Ksf}{t+1}$ MAN multicast messages in an order according to the random permutation, which is unknown by each user, such that each user does not know the $t+1$ effective users for which each MAN multicast message is useful. 

As a result, from the viewpoint of each real user $k\in [\Ksf]$, the compositions of the received multicast messages are equivalent for different demand vectors given $d_k$, such that it cannot get any information about the demands of other real users.

The JCM caching scheme in~\cite{d2dcaching} extends  the $\Ksf$-user MAN caching scheme to the  D2D scenario (without privacy constraint) by using the MAN cache placement and splitting each MAN multicast message in the delivery phase into $t+1$ equal-length  sub-messages, each of which can be transmitted by one of the $t+1$ users.  
 However, it is difficult to use this extension idea to directly extend the above shared-link private caching to the considered D2D private scenario. 
 The main issue is that   we cannot have any virtual transmitter in the system. 
Hence,   instead of  directly extending  the   shared-link private caching scheme proposed in~\cite{wan2019privatecaching}  to the  D2D private model, we will propose a novel and non-trivial D2D private caching scheme.

 
 \section{D2D Caching Schemes with Private  Demands}
\label{sec:general}
A trivial solution is to let  each user   recover the whole library in order to hide its demanded file.  
\begin{thm}[Uncoded Scheme]
\label{thm:scheme 1}
For  the $(\Ksf,\Nsf,\Msf)$ private D2D  caching system, $\Rsf^{\star}$ is upper bounded by   
\begin{align}
\Rsf^{\star}\leq \Rsf_{\text{u}}=\frac{\Ksf}{\Ksf-1}(\Nsf-\Msf). 
\label{eq:scheme 1}
\end{align}
\end{thm}

We then propose a coded private caching scheme with a novel cache placement  based on generating $(\Ksf-1)(\Nsf-1)$ virtual users whose subpacketization is different from the MAN cache placement,  and a novel coded  delivery scheme, the compositions of whose transmitted multicast messages are equivalent from the viewpoint of each real user.  More precisely, from the novel caching construction, the proposed D2D private caching scheme divides the D2D scenario into $\Ksf$ independent shared-link caching models, each of which serves 
\begin{align}
\Usf:= (\Ksf-1)\Nsf \label{eq:def of U}
\end{align}
  effective users. In addition,  instead of assigning one demand to each virtual user in the D2D scenario, we assign one demand to each virtual user for each of the $\Ksf$ divided shared-link models, such that each file is demanded by $\Ksf-1$ effective users to be served in this shared-link model.  
The achieved load is given in the following theorem and the detailed description on the proposed scheme could be found in Section~\ref{sub:main scheme}.
\begin{thm}[Coded Scheme]
\label{thm:main scheme}
For the $(\Ksf,\Nsf,\Msf)$  D2D private caching system, 
$\Rsf^{\star}$ is upper bounded by  the lower convex envelope of 
the following points
\begin{align}
&(\Msf,\Rsf_{\text{c}})= \left( 
\frac{\Nsf+t-1}{\Ksf},
\frac{\binom{\Usf}{t}-\binom{\Usf-\Nsf}{t}}{ \binom{\Usf}{t-1}}
\right), \ \forall  t\in [ \Usf+1].
\label{eq:extended scheme}
\end{align}
\end{thm}
Notice that when $t=\Usf+1$ in~\eqref{eq:extended scheme}, we have the trivial corner point $(\Msf,\Rsf_{\text{c}})=(\Nsf, 0)$.

By comparing the proposed coded private  caching scheme in Theorem~\ref{thm:main scheme} and the converse bound for the shared-link caching problem without privacy constraint in~\cite{yas2}, we have the following order optimality results (whose detailed proof could be found in the extended version of this paper~\cite{optimalprivateD2D}).
\begin{thm}
\label{thm:order optimality}
For  the $(\Ksf,\Nsf,\Msf)$ private D2D caching system, the proposed scheme in Theorem~\ref{thm:main scheme} is order  optimal within a factor of $6$ if $\Nsf\geq \Ksf$  and $\Msf \geq 2\Nsf/\Ksf$, and  a factor of $12$ if $\Nsf<\Ksf $ and $\Msf\geq \Nsf/\Ksf$.
\end{thm}

\begin{rem}
\label{rem:colluding users}
We say that the users in the system {\it collude} if they exchange the indices of their demanded files and their cache contents.  
Privacy constraint against colluding users is a stronger notion than~\eqref{eq:privacy2} and is defined as follows 
\begin{align}
 I\big(\mathbf{D} ; \mathbf{X}, \{Z_k: k\in \Sc \} | \{D_k:k\in \Sc \} \big)=0 , \ \forall \Sc \subseteq [\Ksf],  \Sc\not=\emptyset.   
\label{eq:colluding privacy}
\end{align}
In the extended version of this paper~\cite{optimalprivateD2D}, we also propose a novel converse bound by considering the privacy constraint in~\eqref{eq:colluding privacy}. Comparing the proposed achievable scheme in Theorem~\ref{thm:main scheme} and the novel converse bound, we can prove that the scheme in Theorem~\ref{thm:main scheme} is order optimal  under the constraint of uncoded cache placement and  privacy against colluding users,  within a factor of $18$ (numerical simulations suggest $27/2$).
\end{rem}

\subsection{Example}
\label{sub:example}
Before the general description on the proposed scheme in Theorem~\ref{thm:main scheme}, we first use the following example to illustrate the main idea.
Consider the  $(\Ksf,\Nsf,\Msf)=(2,3,2)$ D2D caching system  with private demands. From~\eqref{eq:extended scheme} and~\eqref{eq:def of U}, in this example we have $t=2$ and $\Usf=3$.

{\it Placement Phase.}
We also use the private placement precoding strategy   proposed in~\cite{wan2019privatecaching}.
Each file $F_i$ where $i\in [\Nsf]$ is divided into $\Ksf\binom{\Usf}{t-1}=6$ non-overlapping and equal-length pieces, denoted by $S_{i,1},\ldots,S_{i,6}$, where each piece has $\Bsf/6$ bits.  
For user $k_1=1$, 
we aim to generate the subfiles for the    shared-link model,  in which user $k_1=1$ broadcast packets and there are $\Ksf-1=1$ real user (user $2$) and $(\Ksf-1)(\Nsf-1)=2$ virtual users (users $3$ and $4$) to  be served. In other words, there are totally $(\Ksf-1)(\Nsf-1)+\Ksf-1=\Usf$   effective users to be served, whose union set is $[(\Ksf-1)(\Nsf-1)+\Ksf]\setminus\{k_1\}=[2:4]$.
We randomly generate a permutation of $\left[ \binom{\Usf}{t-1}\right]=[3]$, denoted by $\pv_{i,k_1}=\pv_{i,1}=(p_{i,1}[1],p_{i,1}[2],p_{i,1}[3])$,  independently and uniformly over the set of all possible permutations. We assume that $\pv_{i,1}=(1,2,3).$ For each set $\Wc\subseteq [(\Ksf-1)(\Nsf-1)+\Ksf]\setminus\{k_1\}=[2:4]$ where $|\Wc|=t-1=1$, we generate a subfile $f^{k_1}_{i,\Wc}$ of $F_i$ which should be cached by users in $\{k_1\}\cup \Wc \cap [\Ksf]$ 
  according to  $\pv_{i,1}$ as follows,
  \begin{subequations}
\begin{align}
&f^1_{i,\{2\}}= S_{i,p_{i,1}[1]}=S_{i,1},   \  f^1_{i,\{3\}}= S_{i,p_{i,1}[2]}=S_{i,2}, \\&   f^1_{i,\{4\}}= S_{i,p_{i,1}[3]}=S_{i,3}.
\end{align}  
\end{subequations}
Hence, $f^1_{i,\{2\}}$ is cached by users $1$ and $2$, while  $f^1_{i,\{3\}}$ and $f^1_{i,\{4\}}$ are only cached by user $1$.
  Similarly, for user $k_2=2$, we randomly generate a permutation of $\left[ \binom{\Usf}{t-1}+1:2\binom{\Usf}{t-1}\right]=[4:6]$, denoted by $\pv_{i,k_2}=\pv_{i,2}=(p_{i,2}[1],p_{i,2}[2],p_{i,2}[3])$,  independently and uniformly over the set of all possible permutations. We assume that $\pv_{i,2}=(4,5,6).$ For each set $\Wc\subseteq [(\Ksf-1)(\Nsf-1)+\Ksf]\setminus\{k_2\}=\{1,3,4\}$ where $|\Wc|=t-1=1$, we generate a subfile $f^{k_2}_{i,\Wc}$ of $F_i$ which should be cached by users in $\{k_2\}\cup \Wc \cap [\Ksf]$ 
  according to  $\pv_{i,2}$ as follows,
  \begin{subequations}
\begin{align}
&f^2_{i,\{1\}}= S_{i,p_{i,2}[1]}=S_{i,4},  \  f^2_{i,\{3\}}= S_{i,p_{i,2}[2]}=S_{i,5},  \\& f^2_{i,\{4\}}= S_{i,p_{i,2}[3]}=S_{i,6}.
\end{align}  
\end{subequations}
Hence, $f^2_{i,\{1\}}$ is cached by users $1$ and $2$, while  $f^2_{i,\{3\}}$ and $f^2_{i,\{4\}}$ are only cached by user $2$.

Recall that $Z_{1}=(\mathscr{M}(C_1),C_1)$ denotes the cache of user $1$. In this example,  
$$
C_1=\cup_{i\in [\Nsf]} \{S_{i,p_{i,1}[1]},S_{i,p_{i,1}[2]},S_{i,p_{i,1}[3]},S_{i,p_{i,2}[1]}\}
$$
 and $\mathscr{M}(C_1)$ denotes the indices of the contained bits in $C_1$.\footnote{\label{foot:no metadata} For sake of simplicity, in the rest of paper, when we describe our achievable scheme, we directly provide $C_k$ or $X_k$ for each user $k\in [\Ksf]$ without repeating that its metadata.} 
Similarly, we can obtain $Z_2$ for user $2$. Hence, 
 each user $k\in [2]$ caches $4 $ pieces of each file, and thus  it totally caches $2\frac{4 \Bsf}{6} =2\Bsf$ bits, satisfying the memory size constraint. 
 In addition, since the random permutations $\pv_{i,1}$ and $\pv_{i,2}$ are unknown to user $k$, each cached subfile of $F_i$ with the same superscript is equivalent from the viewpoint of user $k$, e,g,  $f^1_{i,\{2\}}$, $f^1_{i,\{3\}}$, and $f^1_{i,\{4\}}$ are equivalent from the viewpoint of user $1$.
 Each uncached subfile of $F_i$ with the same superscript is also  equivalent from the viewpoint of user $k$, e.g.,  $f^2_{i,\{3\}}$ and  $f^2_{i,\{4\}}$ are equivalent from the viewpoint of user $1$.

We will consider two demand vectors $(1,2)$, $(1,1)$, which represent all possible non-equivalent demand configurations. %

{\it Delivery Phase for $\dv=(1,2)$.}
We treat the $\Ksf$ transmissions from the $\Ksf$ users as $\Ksf$ shared-link transmissions.

Let us first consider the  $1^{\text{st}}$ shared-link transmission  in which user $k_1=1$ broadcast packets.
We  assign one demanded file to each virtual user such that each file in the library is demanded by $\Ksf-1=1$ effective user in $[2:4]$.
More precisely, we first let $d^{1}_2=d_2=2$, representing the demanded file  by real user $2$  in the $1^{\text{st}}$  shared-link transmission.
 We also  let $d^{1}_{3}=1$ and $d^{1}_{4}=3$, representing the demanded files by virtual users $3$ and $4$ in the $1^{\text{st}}$  shared-link transmission, respectively. 
For each set $\Sc \subseteq [(\Ksf-1)(\Nsf-1)+\Ksf]\setminus \{k_1\}=[2:4]$ where $|\Sc|=t=2$, we generate 
\begin{align}
W^{k_1}_{\Sc}=\underset{j\in \Sc}{\oplus } f^{k_1}_{ d^{k_1}_{j}, \Sc\setminus \{j\}}.\label{eq:proposed multicast message}
\end{align}
In this example, we have 
\begin{subequations}
\begin{align}
&W^{1}_{\{2,3\}}=f^1_{2,\{3\}}\oplus f^1_{1,\{2\}}, \  \ W^{1}_{\{2,4\}}=f^1_{2,\{4\}}\oplus f^1_{3,\{2\}}, \\ & W^{1}_{\{3,4\}}=f^1_{1,\{4\}}\oplus f^1_{3,\{3\}}.
\end{align}
\label{eq:three messages}
\end{subequations}
Finally, we generate one permutation of $\left[\binom{\Usf}{t} \right]=[3]$, denoted by $\qv_{k_1}=\qv_1=(q_{1,1},q_{1,2},q_{1,3})$, independently and uniformly over the set of all possible permutations. 
By assuming $\qv_{1}=(1,2,3)$ which is used to transmit the three multicast messages in~\eqref{eq:three messages} in an order which is unknown to users, we can hide the users to whom each multicast message is useful. Hence, we let
$P_{k_1}=P_1=(W^{1}_{\{2,3\}},W^{1}_{\{2,4\}},W^{1}_{\{3,4\}})$.
The trusted server transmits $\mathscr{M}(P_1)$ to user $1$, who is then instructed to broadcast $X_1=(\mathscr{M}(P_1), P_1)$.   

We then consider the $2^{\text{nd}}$ shared-link transmission, in which user $k_2=2$ broadcast packets as the server.
Similarly, we   let $d^{2}_1=d_1=1$,  $d^{2}_{3}=2$, $d^{2}_{4}=3$, and generate 
\begin{subequations}
\begin{align}
&W^{2}_{\{1,3\}}=f^2_{1,\{3\}}\oplus f^2_{2,\{1\}},  \  W^{2}_{\{1,4\}}=f^2_{1,\{4\}}\oplus f^2_{3,\{1\}},   \\ & W^{2}_{\{3,4\}}=f^2_{2,\{4\}}\oplus f^2_{3,\{3\}}.
\end{align}
\label{eq:three messages 2}
\end{subequations}
By generating a random permutation of $[3]$, denoted by  $\qv_2$ (assumed to be $(1,2,3)$), we let $P_{2}=(W^{2}_{\{1,3\}}, W^{2}_{\{1,4\}}, W^{2}_{\{3,4\}})$. The trusted server transmits $\mathscr{M}(P_2)$ to user $2$, who is then instructed to broadcast  $X_2=(\mathscr{M}(P_2), P_2)$.
From the received packets, each user $k\in[2]$ can recover its demanded file. For the privacy, we then focus on the demand vector $\dv=(1,1)$.

{\it Delivery Phase for $\dv=(1,1)$.}
By the same method as described above,  
user $1$ is instructed to broadcast $ X_1=(\mathscr{M}(P_1), P_1)$ where $P_1=(W^{1}_{\{2,3\}},W^{1}_{\{3,4\}},W^{1}_{\{2,4\}})$\footnote{\label{foot:order}The order of the multicast messages in $P_1$ is not important because this order is generated randomly. Here, we assume this order for sake of easy comparison with the demand vector $(1,2)$.} and
\begin{subequations}
\begin{align}
&W^{1}_{\{2,3\}}=f^1_{1,\{3\}}\oplus f^1_{2,\{2\}}, \  \ W^{1}_{\{3,4\}}=f^1_{2,\{4\}}\oplus f^1_{3,\{3\}},   \\ & W^{1}_{\{2,4\}}=f^1_{1,\{4\}}\oplus f^1_{3,\{2\}}.
\end{align}
 \label{eq:second demand W1}
\end{subequations}
User $2$ is instructed to broadcast  $ X_2=(\mathscr{M}(P_2), P_2)$ where $P_2=(W^{2}_{\{1,3\}},W^{2}_{\{1,4\}},W^{2}_{\{3,4\}})$ and
\begin{subequations}
\begin{align}
&W^{2}_{\{1,3\}}=f^2_{1,\{3\}}\oplus f^2_{2,\{1\}},  \ \ W^{2}_{\{1,4\}}=f^2_{1,\{4\}}\oplus f^2_{3,\{1\}},   \\ & W^{2}_{\{3,4\}}=f^2_{2,\{4\}}\oplus f^2_{3,\{3\}}.
\end{align}
 \label{eq:second demand W2}
\end{subequations}

{\it Privacy.}
Let us focus on user $1$. For each demand vector, the delivery scheme is equivalent to two independent shared-link transmissions, and in the $k^{\text{th}}$ shared-link transmission where $k\in [2]$ only the subfiles with superscript $k$ are transmitted by user $k$. In other words, no subfile appears in the two shared-link transmissions simultaneously. Each shared-link transmission is equivalent to a shared-link private caching scheme  in~\cite{wan2019privatecaching} where each file in the library is demanded by $\Ksf-1=1$ effective users. By the construction on the cache placement, each cached subfile of $F_i$ with the same superscript is equivalent from the viewpoint of each real user, while
 each uncached subfile of $F_i$ with the same superscript is also  equivalent from the viewpoint of this user. Hence, the $k^{\text{th}}$ shared-link transmissions for different demand vectors are equivalent from the viewpoint of each real user. For example, for user $1$, $f^1_{1,\{2\}}$ and $f^1_{1,\{3\}}$ are equivalent, 
while $f^1_{2,\{3\}}$ and $ f^1_{2,\{2\}}$ are equivalent. Hence, $f^1_{2,\{3\}}\oplus f^1_{1,\{2\}}$ transmitted for demand $(1,2)$ is  equivalent to $f^1_{1,\{3\}}\oplus f^1_{2,\{2\}}$ transmitted for demand $(1,1)$ from the viewpoint of user $1$. Similarly, $X_1$ transmitted for     demand $(1,2)$ is  equivalent to $X_1$ transmitted for demand $(1,1)$ from the viewpoint of user $1$.
By the same reasoning, it can be checked that $X_2$'s transmitted for different demands are also equivalent  from the viewpoint of user $1$. In conclusion, user $1$ does not know any information about the demand of user $2$ from the transmission. 

Similarly, it can be seen that the privacy of the demand of user $1$ is also preserved from user $2$. Hence, the proposed D2D coded private caching scheme is indeed private.

 {\it Performance.}
Each user transmits three binary sums of subfiles, each of which has $\Bsf/6$ bits. Hence, the achieved load is $1$, while the load achieved by the uncoded scheme in Theorem~\ref{thm:scheme 1} is $2$ and the JCM caching scheme without privacy achieves $2/3$.

\subsection{Proof of Theorem~\ref{thm:main scheme}}
\label{sub:main scheme}
We are now ready to generalize the example in Section~\ref{sub:example}. Recall $\Usf= (\Ksf-1)\Nsf$ defined in~\eqref{eq:def of U}. We focus on the memory size $\frac{(\Ksf-1)(t-1)+\Usf }{\Ksf\Usf}\Nsf$, where $t\in [ \Usf]$.  
We generate $(\Ksf-1)(\Nsf-1) $ virtual users, which are labelled as users $\Ksf+1,\ldots, (\Ksf-1)(\Nsf-1)+\Ksf$. 

 {\it Placement Phase.}
Each file $F_i$ where $i\in [\Nsf]$ is divided into $\Ksf\binom{\Usf}{t-1}$ non-overlapping and equal-length pieces, denoted by $S_{i,1},\ldots,S_{i,\Ksf\binom{\Usf}{t-1}}$, where each piece has $\frac{\Bsf}{\Ksf\binom{\Usf}{t-1}}$ bits.  
For each user $k\in [\Ksf]$, 
we aim to generate the subfiles for the $k^{\text{th}}$  shared-link model,  in which user $k$ broadcast packets as the server and there are $\Ksf-1=1$ real user   and $(\Ksf-1)(\Nsf-1)=2$ virtual users to  be served. In other words, there are totally $(\Ksf-1)(\Nsf-1)+\Ksf-1=\Usf$   effective users to be served, whose union set is  $[(\Ksf-1)(\Nsf-1)+\Ksf]\setminus\{k\}$.
We randomly generate a permutation of $\left[ (k-1)\binom{\Usf}{t-1}+1: k\binom{\Usf}{t-1} \right] $, denoted by $\pv_{i,k}=\left(p_{i,k}[1],\ldots,p_{i,k}\left[ \binom{\Usf}{t-1}\right] \right)$,  independently and uniformly over the set of all possible permutations. 
We sort all sets $\Wc\subseteq  [(\Ksf-1)(\Nsf-1)+\Ksf]\setminus\{k\}$ where $|\Wc|=t-1$,  in a lexicographic order, denoted by $\Wc(1),\dots, \Wc\left(\binom{\Usf}{t-1} \right)$.
For each $j\in \left[\binom{\Usf}{t-1} \right]$, we   generate a subfile  
\begin{align}
f^k_{i, \Wc(j)}= S_{i,p_{i,k}[j]},\label{eq:extend assignement 1}
\end{align}
which is cached by users in $\{k\}\cup  \Wc(j) \cap [\Ksf]$.

After considering all $\Ksf$ shared-link models, 
 each real user $k \in [\Ksf]$  caches  all $\binom{\Usf}{t-1} $ subfiles with superscript $k$, and $\binom{\Usf-1}{t-2}$ subfiles with superscript $k^{\prime}$ for each $k^{\prime} \in [\Ksf]\setminus \{k\}$. 
 Hence, user $k$ totally caches $\binom{\Usf}{t-1}+(\Ksf-1)\binom{\Usf-1}{t-2}$ subfiles, each of which has $\frac{\Bsf}{\Ksf\binom{\Usf}{t-1}}$ bits, and thus the number of cached bits is $\frac{\binom{\Usf}{t-1}+(\Ksf-1)\binom{\Usf-1}{t-2}}{\Ksf\binom{\Usf}{t-1}}\Bsf=\Msf\Bsf$.
Moreover, for each file $i\in [\Nsf]$, the random permutations $\pv_{i,j}$ where $j\in [\Ksf]$ are unknown to user $k\in [\Ksf]$.   Hence,  from the viewpoint of user $k$, each cached subfile of $F_i$ with the same superscript is equivalent from the viewpoint of user $k$, while each uncached subfile of $F_i$ with the same superscript is also equivalent.

 {\it Delivery Phase.} 
We divide the transmissions from the $\Ksf$ into $\Ksf$ shared-link transmissions. Let us  focus on the $k^{\text{th}}$ shared-link transmission, where $k\in [\Ksf]$.
 
We first assign one demanded file to each virtual user such that each file in the library is demanded by $\Ksf-1$ effective user.
More precisely,   for each real user $k^{\prime}\in [\Ksf]\setminus\{k\}$, let 
\begin{align}
 d^k_{k^{\prime}}=d_{k^{\prime}}. \label{eq:real user demand}
\end{align}
We then define 
\begin{align}
 n_{i,k}:=|\{k^{\prime}\in[\Ksf]\setminus\{k\}:d_{k^{\prime}}=i \}|, \ \forall i\in [\Nsf],
\end{align} 
which represents the number of real users in $[\Ksf]\setminus \{k\}$ demanding $F_i$. 
One file is assigned to each of the $(\Ksf-1)(\Nsf-1)$ virtual users as follows. 
For each file $i\in [\Nsf]$, we let 
\begin{align}
 d^k_{1+\Ksf+(i-1)(\Ksf-1)-\sum_{q\in [i-1]}n_{q,k}}=\dots=d^k_{\Ksf+i(\Ksf-1)-\sum_{q\in[i] }n_{q,k}}=i.
 \label{eq:demand assignment}
\end{align}
For example, when $i=1$, we let  
\begin{align*}
d^k_{\Ksf+1}=\dots=d^k_{2\Ksf-n_{1,k}-1}=1,
\end{align*}
when $i=2$, we let 
\begin{align*}
d^k_{2\Ksf-n_{1,k}}=\dots=d^k_{3\Ksf-n_{1,k}-n_{2,k}-2}=2,
\end{align*}
and so on. 
Hence, each file is requested by $\Ksf-1$ effective users in the user set  $[(\Ksf-1)(\Nsf-1)+\Ksf]\setminus\{k\}$.
For each file, we randomly and uniformly choose an effective user demanding this file as a leader user. The leader set is denoted by $\Lc_k$.
 
We generate a  random permutation of   $[(\Ksf-1)(\Nsf-1)+\Ksf]\setminus \{k\}$, denoted by $\qv_{k}=\left(q_{k,1},\ldots,q_{k,\Usf} \right)$, independently and uniformly over the set of all possible permutations.

For each set $\Sc \subseteq [\Usf]$ where $|\Sc|=t$, by computing $\Sc^{\prime}= \cup_{j^{\prime}\in \Sc} \{q_{k,j^{\prime}}\}$, we generate the multicast message  
$
 W^{k}_{\Sc}=\underset{j\in \Sc}{\oplus } f^{k}_{ d^{k}_{q_{k,j}}, \Sc^{\prime} \setminus \{q_{k,j}\}}  
$ as in~\eqref{eq:proposed multicast message}.
The trusted server asks  user $k$  to broadcast $ X_k=(\mathscr{M}(P_k), P_k)$ to other users, where
\begin{align}
 P_{k}=\left( W^{k}_{\Sc}: (\cup_{j^{\prime}\in \Sc} \{q_{k,j^{\prime}}\}) \cap \Lc \neq \emptyset \right),\label{eq:X_k}
\end{align}
Notice that in the metadata of $W^{k}_{\Sc}$, the set $\Sc$ is revealed.
 
{\it Decodability.}
We focus on user $k\in [\Ksf]$.
In the $j^{\text{th}}$ transmission where $j\in [\Ksf]\setminus \{k\}$, it was shown in~\cite[Lemma 1]{exactrateuncoded}, user $k$ can reconstruct each multicast message   $W^{j}_{\Sc}$ where $\Sc \subseteq [\Usf]$ and $|\Sc|=t$.
User $k $ then checks each  $W^{j}_{\Sc}$  where $\Sc \subseteq [\Usf]$ and $|\Sc|=t$. If $W^{j}_{\Sc}$  contains $t-1$ cached subfiles  and one uncached subfile, user $k$ knows this message is useful to it and decodes the uncached subfile. 
  
It is obvious that each   subfile of $F_{d_k}$ which is not cached by user $k$, appears in one multicast message.  Hence, after considering all transmitted packets in the delivery phase, user $k\in [\Ksf]$ can recover all requested subfiles to reconstruct its requested file.

{\it Privacy.}
The intuition on the privacy is the same as the above example and
   the information-theoretic proof on the privacy can be found in~\cite{optimalprivateD2D}.

 {\it Performance.}
Each user $k\in [\Ksf]$ broadcasts $\binom{\Usf}{t}-\binom{\Usf-\Nsf}{t}$ multicast  messages, each of which contains  $\frac{\Bsf}{\Ksf\binom{\Usf}{t-1}}$ bits. Hence, the achieved  load  coincides with~\eqref{eq:extended scheme}.

 \section{Numerical Evaluations}
We   provide numerical evaluations of the proposed private caching schemes for the $(\Ksf,\Nsf,\Msf )=(10,5,\Msf)$ D2D caching system with private   demands. We  
compare the baseline D2D uncoded private caching scheme in Theorem~\ref{thm:scheme 1} and the coded caching scheme in Theorem~\ref{thm:main scheme}, with 
  the converse bound in~\cite{yas2} for the shared-link caching model.
It shows that 
the proposed coded caching scheme  outperforms  the uncoded scheme.

\begin{figure}
\centerline{\includegraphics[scale=0.45]{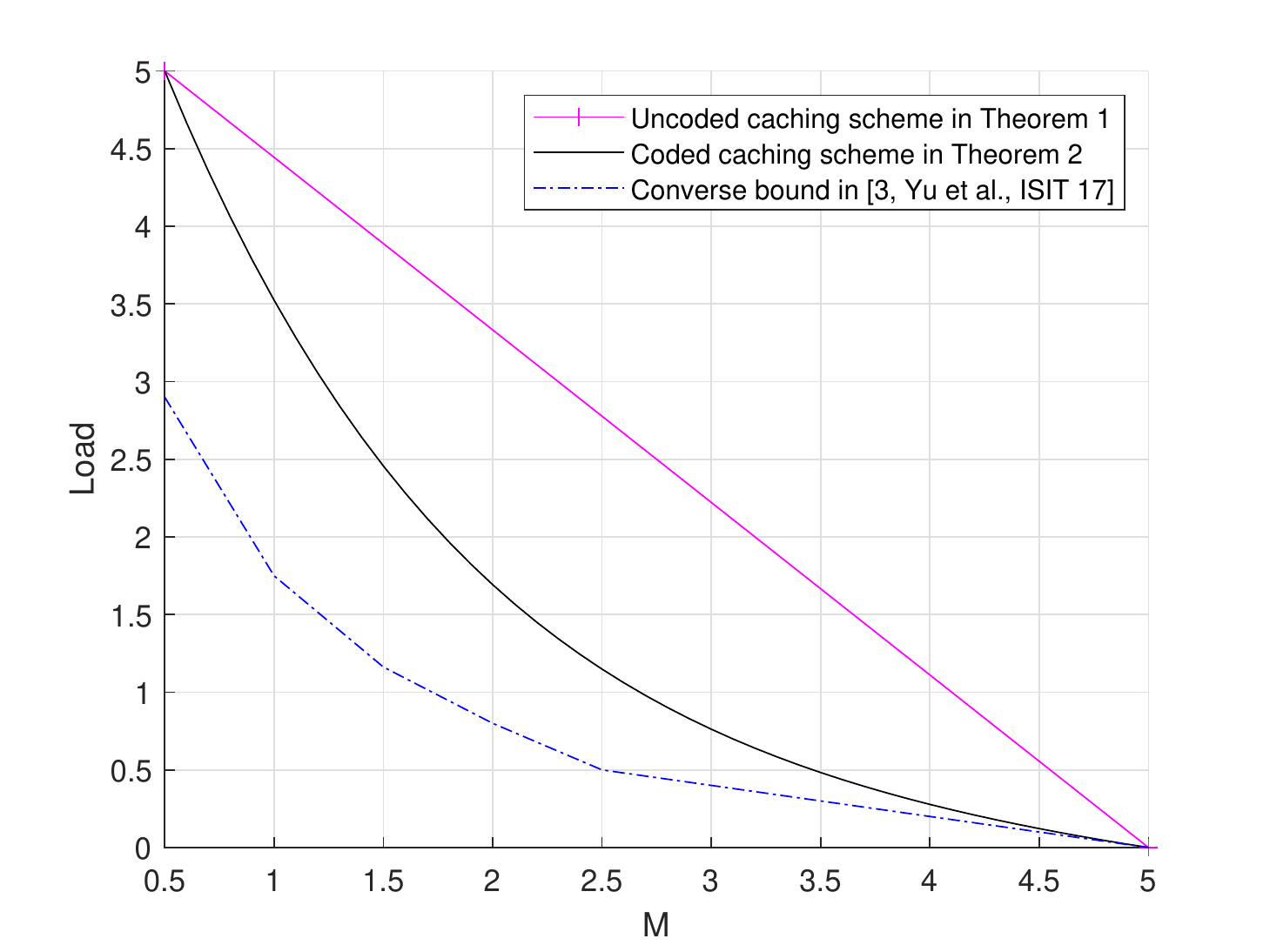}}
\caption{\small $(\Msf, \Rsf)$ tradeoff for the $(\Ksf,\Nsf,\Msf)=(10,5,\Msf)$ D2D caching system with private  demands.}
\label{fig:num}
\vspace{-5mm}
\end{figure}

\section{Conclusions}
\label{sec:conclusion} 
We introduced a novel D2D private caching model with a trusted server, which aims to preserve the privacy of the users demands. 
We proposed  a novel D2D private coded caching scheme, which is order optimal   within a factor of $6$ when $\Nsf\geq \Ksf$ and $\Msf\geq 2\Nsf/\Ksf$, and within a factor of $12$ when $\Nsf<\Ksf$ and $\Msf\geq \Nsf/\Ksf$.
This scheme is also order optimal within a factor of $18$ for any system parameters  under the constraint of uncoded cache placement and  privacy against colluding users. 
 
\section*{Acknowledgemnt} 
The work of K. Wan and G. Caire is supported by  the European Research Council  under the ERC Advanced Grant N. 789190, CARENET. The work of H. Sun was supported in part by NSF Award 2007108. 
The work of D. Tuninetti is supported in part by the National Science Foundation Award 1910309.
The work of M. Ji is supported in part by NSF Awards 1817154 and 1824558.

\bibliographystyle{IEEEtran}
\bibliography{IEEEabrv,IEEEexample}

\end{document}